\documentclass[aps,reprint,prb,showpacs,showkeys,preprintnumbers,amsmath,amssymb,citeautoscript,floatfix,superscriptaddress]{revtex4-1}
\usepackage{graphicx}
\usepackage{pstricks}
\usepackage{url}

\usepackage{breakurl}
\usepackage[breaklinks]{hyperref}
\usepackage{subfigure}
\usepackage{dcolumn}
\usepackage{bm}

\usepackage{mathptmx}
\usepackage{import}
\usepackage[toc,page]{appendix}
\usepackage{chemformula}
\usepackage{amsmath}
\usepackage{placeins}

\newcounter{saveeqn}

\begin{document}

\title{Recurrent Neural Networks for Prediction of Electronic Excitation Dynamics}

\author{Ethan P. Shapera}
\email{epshaper@buffalo.edu}
\affiliation{Department of Chemistry, State University of New York at Buffalo, Buffalo, USA}
\affiliation{Institute of Theoretical Physics and Computational Physics, Graz University of Technology, 8010 Graz, Austria}

\author{Cheng-Wei Lee}
\affiliation{Colorado School of Mines, Golden, CO 80401, USA}


\date{\today}

\begin{abstract}
We demonstrate a machine learning based approach which can learn the time-dependent electronic excitation dynamics of small molecules subjected to ion irradiation.
Ensembles of recurrent neural networks are trained on data generated by time-dependent density functional theory to relate atomic positions to occupations of molecular orbitals.
New data is incrementally and efficiently added to the training data using an active learning process, thereby improving model accuracy.
Predicted changes in orbital occupations made by the recurrent neural network ensemble are found to have errors and one standard deviation uncertainties which are two orders of magnitude smaller than the typical values of the orbital occupation numbers.
The trained recurrent neural network ensembles demonstrate a limited ability to generalize to molecules not used to train the models.
In such cases, the models are able to identify key qualitative features, but struggle to match the quantitative values.
The machine learning procedure developed here is potentially broadly applicable and has the potential to enable study of broad ranges of both materials and dynamical processes by drastically lowering the computational cost and providing surrogate model for multiscale simulations.
\
\end{abstract}

\maketitle

Machine learning (ML) has been an immensely successful and rapidly developing tool for materials science because it enables both analyzing large volumes of data to identify trends and designing new material \cite{pilania2021}.
Diverse approaches to ML have been applied to diverse topics including: crystal structure prediction\cite{egorova2020,wengert2021,cheng2022}, electronic stopping power \cite{ward2023}, metamaterials\cite{kennedy2022,ha2023}, microscopy \cite{waller2015,kalinin2022}, superconductivity\cite{roter2020,yazdani2023}, and force fields for molecular dynamics\cite{botu2017,unke2021}, among many others.
However, the vast majority of this ML for application to DFT has focused on time-independent properties; few studies have sought to predict time-dependent system dynamics.
Among the limited examples, Meinecke \textit{et. al}\cite{meinecke2023} demonstrated the prediction of electron-phonon coupling dynamics in a model of MoSe$_2$ using an autoregressive machine learning algorithm, while Shah \textit{et. al}\cite{shah2024} employed neural operators in order to learn propagators for describing the dynamics of electrons subjected to optical excitation.
Further development of machine learning approaches for studying time-dependent phenomena has the potential to accelerate development in two key challenges\cite{hemminger2007} for materials science: controlling materials processes at the level of electrons and controlling matter away from equilibrium.

In this letter, we will demonstrate an ML workflow which produces highly accurate predictions of electronic excitations, consistent with real-time  time-dependent density functional theory (rt-TDDFT), at low computational cost.
Our ML approach for predicting the electronic response has two key components: the responses are modeled using an ensemble of recurrent neural networks (RNNs) and the use of active learning to select new data for incorporation into the training data.
RNNs are a variant of neural network which include loops between nodes within the network, preserving a memory of previous inputs.
The ability to remember previous states makes RNNs well suited for modelling time-dependent processes.\cite{tealab2018} 
The active learning process provides a systematic method for selecting new data to incorporate into model training.
When combined, these techniques produce ML models which consistently reproduce the results of expensive calculations, but with a computational cost reduced by multiple orders of magnitude.

As a test case, we consider the real-time electronic response of small molecules (CH$_x$Cl$_y$ with $x$+$y$ = 4) under ionizing proton irradiation. 
In these simulations, a proton is launched with a range of speeds and trajectories at a stationary molecule of CH$_x$Cl$_y$.
Occupations of the molecular orbitals then evolve in time following the time-dependent Kohn-Sham equation \cite{runge1984} as a consequence of the time-dependent external potential introduced by the moving proton.
Thus far, accurately simulating the dynamics of such processes requires the use of rt-TDDFT simulations.\cite{Kononov_MRScomm_2022,Sun_PRL_2021,Xu_JACS_2024}
However, rt-TDDFT simulations are often limited to a few snapshots of the statistical ensemble of target molecules and a limited set of interaction impact parameters due to high computational costs.\cite{ward2023, Lee_PRB_2020,Kononov_npj_CM_2023,Shepar_PRL_2023} 
The RNN ensemble predictions are found to reproduce key qualitative features in the excitation dynamics with with model errors and uncertainties two orders of magnitude smaller than typical predicted quantities. The trained model can then efficiently and accurately sample different impact parameters and achieve full statistical sampling. 
In short, our machine learning scheme can enable further study of time-dependent phenomena which have thus far been limited if not inaccessible due to the high computational cost of rt-TDDFT calculations.

\begin{figure}
\includegraphics[width=0.9\linewidth]{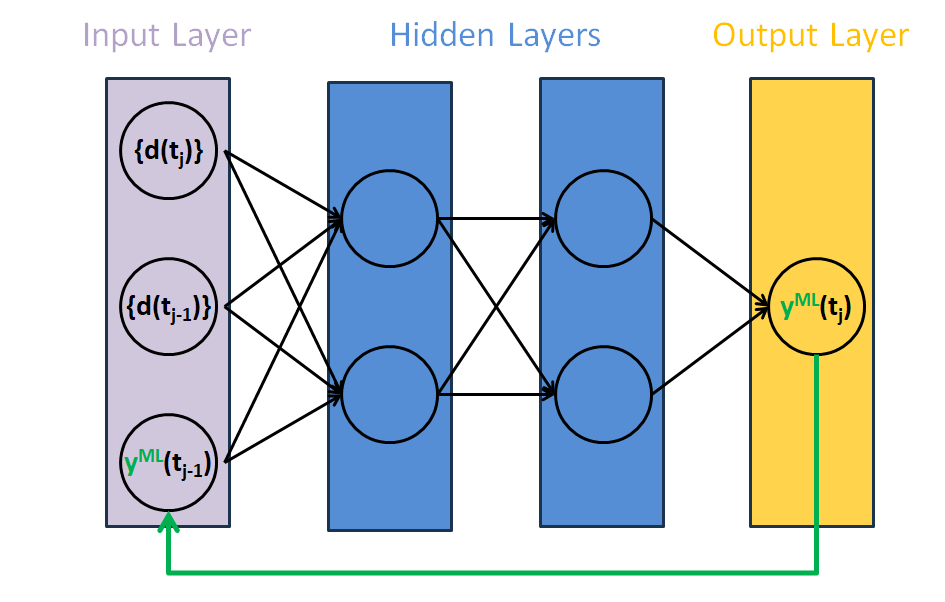}
\caption{\label{fig:RNN}
Schematic of the RNN architecture used in this work.
The green text emphasize the feedback of the target values from output at time step $t_j$ to input at timestep $t_{j+1}$.
}
\end{figure}

The RNNs used in this work were based on the multilayer percepton (MLP) regressors available in \texttt{scikit-learn}\cite{scikit-learn}.
A schematic diagram is provided in Figure \ \ref{fig:RNN}.
The feed-forward MLP regressor was converted into an RNN by feeding the predicted target value from the previous timestep, $y^{ML}(t_{j-1})$,back into the network as a descriptor for the system at timestep $t_{j}$, represented as the green arrow in Figure \ \ref{fig:RNN} which transmits the result from the output layer into the input layer.
Further, the atomic configuration descriptors $\left\{d(t_{j-1})\right\}$ from timestep $t_{j-1}$ were also included as descriptors at timestep $t_{j}$.
Neural networks consisted of three deep layers with 64 nodes in each layer.
All nodes had rectified linear unit (ReLU) activation functions.
Network weights and biases were optimized using a stochastic gradient-based optimizer \cite{kingma2014} with a fixed learning rate of 0.001.
The configuration of each system at every timestep was featurized using the descriptor sets presented in Ref. \ \onlinecite{shapera2024}.
These descriptors were the crystal graph singular values (CGSVs), Coulomb matrix eigenvalues, and simulation region shape parameters.
This choice of descriptors has been shown to produce to machine learning models with similar performance to the crystal graph representation by Xie \textit{et. al.}\cite{xie2018}, but at reduced computational cost due to a reduction of the dimensionality of the representation by multiple orders of magnitude.
Two target quantities were selected for building machine learning models: change in total occupation number of the valence orbitals ($O$) and change in total energy of the target molecule ($E$).
Changes in occupation number and total energy were used so that values fall in similar ranges between different target molecules.

\begin{figure}
\includegraphics[width=0.9\linewidth]{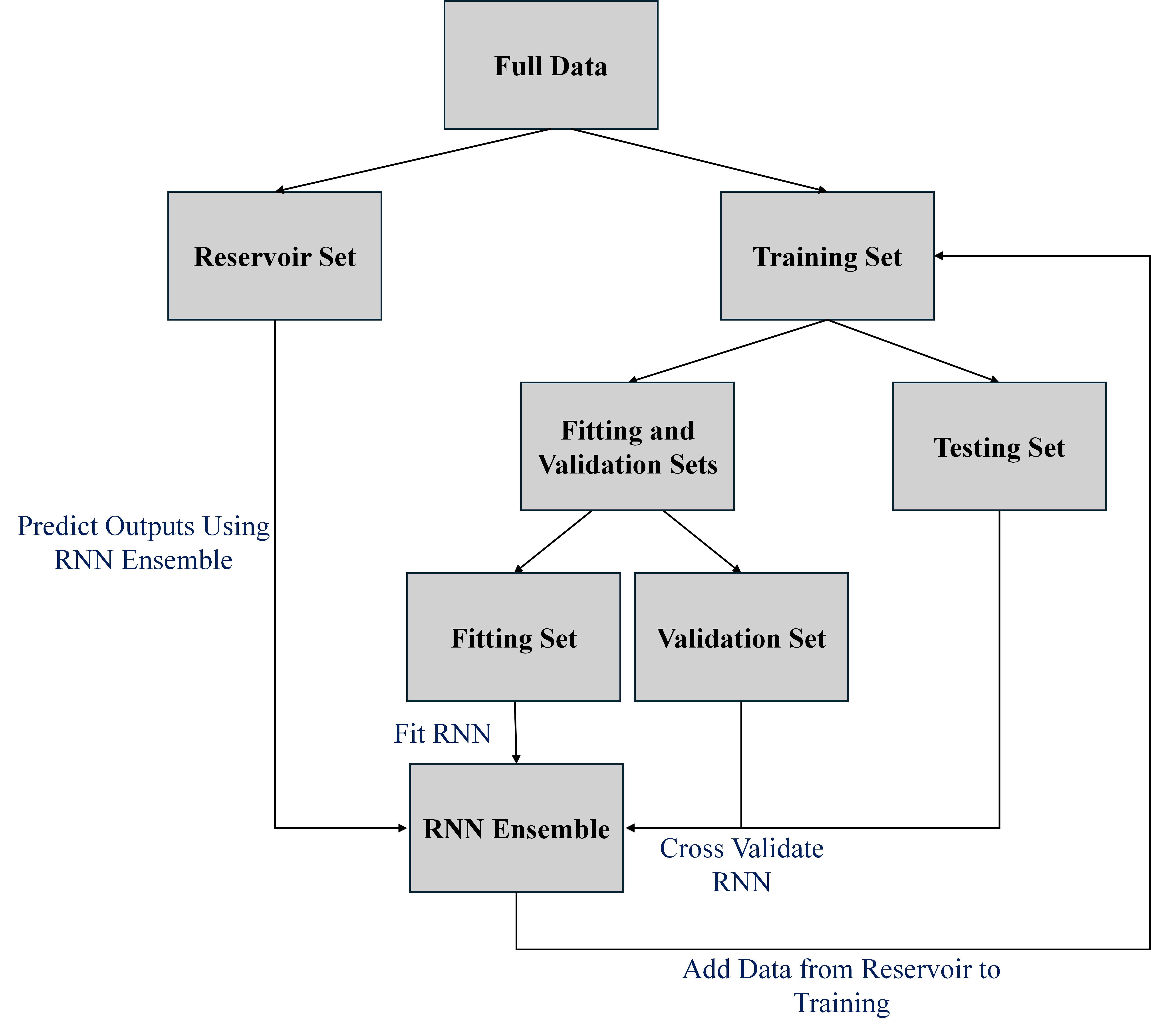}
\caption{\label{fig:ActiveFlowchart}
Flowchart of the active learning and model validation processes used to train RNN ensembles.
}
\end{figure}

Data was selected for inclusion in the training data for the RNN models using an active learning scheme.
A flowchart of the active learning process, combined with the model validation process, is presented in Figure \ \ref{fig:ActiveFlowchart}.
Throughout all model building, each simulation would be left as a single piece and not split between different data sets.
In the $0^{th}$ active learning iteration, the training set was initialized with 10 simulations corresponding to near-hit trajectories with varying speeds by the proton on the central C atom for each target molecule.
All other simulations were assigned to a reservoir set.

Within each active learning iteration, a random 10\% of simulations in the training set was assigned to the testing set and the remaining 90\% assigned to a joint fitting and validation set.
The joint fitting and validation set was randomly partitioned into 10 different fitting and validation sets.
Partitions were performed such that each fitting set contained 80\% of the simulations in the training set and each validation set contained 10\%.
Each fitting set was used to fit an RNN, thereby forming an ensemble of 10 RNNs.
Once fit, each RNN was used to predict the target quantities in the corresponding validation set and the testing set for the active learning iteration.
Performance of each RNN was quantified using the mean absolute error (MAE) between the TDDFT-calculated target quantity and the RNN predicted value.
Once fit and validated, each RNN was used to predict the target values for all runs in the reservoir set.
The 10 simulations in the reservoir set for which the RNN ensemble produced the largest MAEs were moved to the training set in the next active learning iteration.

We report the results from validation tests for RNN ensembles used to predict the change in total valence orbital occupation during the molecule-proton interaction.
Similar results for predicting the total energy during the irradiation process are provided in the Supplemental Materials. 
Figure \ \ref{fig:activeMAE-occ}a plots the MAE values of an RNN ensemble used to predict the change in valence orbital occupation over 15 iterations of active learning.
The MAE values for the fitting, validation, and testing sets decrease non-monotonically and reach a minimum at the 11$^{\mathrm{th}}$ active learning iteration.
After the 11$^{\mathrm{th}}$ active learning iteration, MAE values increase in each successive iteration.
This behavior may be a consequence of adding more data to the training set.
With more data used to fit the RNN ensemble, each RNN becomes less overparameterized, eventually leading the RNNs to struggle to fit data as well\cite{allen2019,chen2019}.

Figure \ \ref{fig:activeMAE-occ}b shows the distribution of MAE values for each run used in the fitting and validation sets for fitting each RNN in the 11$^{\mathrm{th}}$ active learning iteration.
The 11$^{\mathrm{th}}$ iteration is chosen because the MAE values reach a minimum at this iteration.
Here the MAE values are 0.023 e$^-$ for the fitting and validation sets and 0.027 e$^-$ for the testing set.
The matching MAE values for fitting and validation, along with the close match of the MAE distributions demonstrates that the model has not been overfit.

We provide a representative example of the performance of the RNN ensemble in the 11$^{\mathrm{th}}$ active learning iteration for predicting the change in occupation during an irradiation process in Figure \ \ref{fig:activeMAE-occ}c.
In the representative example, a proton projectile is launched at a CH$_4$ molecule with a speed of 1.0 a.u. ($\approx$ 2.2 $\times$ 10$^6$ m s$^{-1}$).
The projectile reaches a minimum distance from the C-atom of 0.4 a$_B$ ($\approx$ 21 pm).
This specific run appeared in the reservoir set when training the RNN ensemble.
The RNN ensemble produced remarkable quantitative and qualitative agreement with the rt-TDDFT calculation.
The RNN ensemble has MAE of 0.02 e$^-$ averaged over the full run with a one standard deviation (1-$\sigma$) uncertainty of 0.07 e$^-$, compared against a final change in occupation of -2.31 e$^-$.
The MAE and 1-$\sigma$ uncertainty are two orders of magnitude smaller than the typical final change in valence orbital occupation.

The RNN ensemble correctly predicts three distinct behavior regimes.
For times less than $\approx$15 a.u. ($\approx$ 0.36 fs), orbital occupations were unchanged because the projectile was not close enough to the target to cause excitations.
Near 20 a.u., there was a rapid change in occupation number as the projectile made the closest approach to the target molecule.
For times greater than $\approx$30 a.u., the total valence orbital occupation underwent slow, low-amplitude oscillations.

Over the 15 active learning iterations, fitting the RNN ensembles required an average of 1.9 core-hours on a workstation per active learning iteration (each iteration consisting of one ensemble of 10 RNNs).
In the selected 11$^{\mathrm{th}}$ active learning iteration, the RNN ensemble was trained in 1.4 core-hours using a total of 104 TDDFT runs.
A single TDDFT run required approximately 300 core-hours on a computer cluster for a single fully converged calculation.
For both the occupation and energy models, running a single TDDFT calculation is approximately two orders of magnitude more computationally expensive than fitting one RNN ensemble.
Once enough TDDFT runs have been performed to train an RNN ensemble using the active learning process, it is more cost effective to train then apply the RNN ensemble approach than to run even a single additional rt-TDDFT calculation.
The reduction in computational cost compared to running further rt-TDDFT calculations is particularly noteworthy here because the target CH$_x$Cl$_y$ is a low cost calculation with few valence electrons and an adiabatic semilocal exchange-correlation (XC) functional.
Larger target materials modeled with advanced XC functionals like HSE06 will face generally two orders of magnitude increase in computational costs,\cite{yk_2021_jcp,Kononov_MRScomm_2022} thereby making the RNN ensemble approach more beneficial as the explicit calculations become more resource intensive.

\begin{figure}
\includegraphics[width=0.8\linewidth]{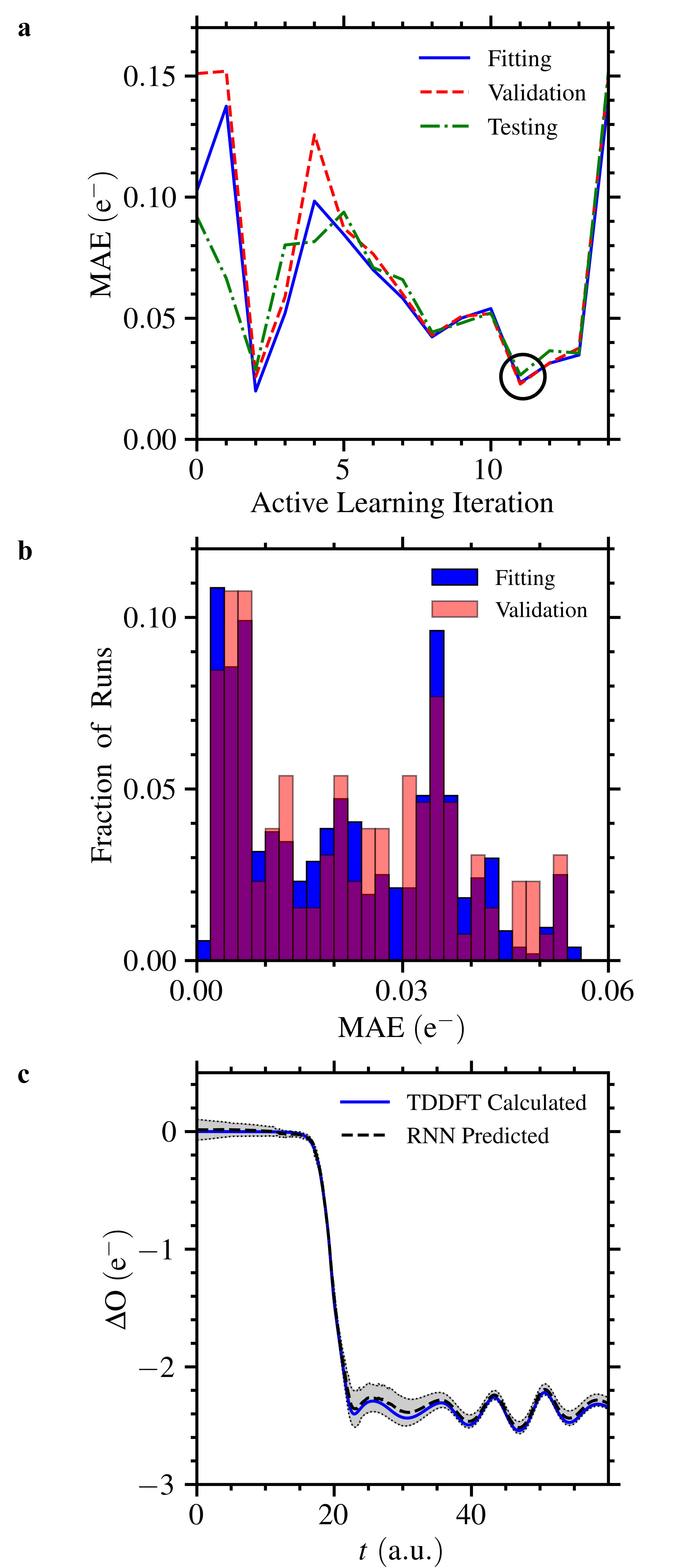}
\caption{\label{fig:activeMAE-occ}
a. Variation in MAE of valence orbital occupation number for ensemble of 10 RNNs trained on CH$_4$ and CHCl$_3$ as more data is added via active learning.
The 11$^{\mathrm{th}}$ iteration, which minimizes the MAEs and is the focus of further discussion, is circled.
b. Distribution of run MAE values in the 11$^{\mathrm{th}}$ active learning iteration for all runs in the fitting and validation sets for the ensemble of 10 RNNs.
c.Representative example of RNN ensemble prediction results for the change in total occupation of the valence orbitals.
A proton projectile with a speed of 1.0 a.u. passes within 0.4 a$_B$ ($\approx$ 21 pm) of the C-atom in CH$_4$.
Gray shaded regions show the 1$\sigma$ uncertainty of the RNN ensemble predictions.
}
\end{figure}

In this work, we have demonstrated the use of ensembles of RNNs to predict the dynamical behavior of small molecules under ion irradiation, along with an estimated uncertainty.
RNN ensembles were trained on rt-TDDFT data simulating  CH$_4$ and CHCl$_3$ irradiated by ionizing proton.
Two time-dependent quantities were considered: the total occupation of valence orbitals in the target molecule and the total energy of the molecule system.
The models showed remarkable accuracy and precision in reproducing the calculated time-dependent behavior.
Models consistently reproduced key features of the dynamics and showed close agreement with rt-TDDFT results for fitting, validation, and testing data sets.
The models were unable to generalize to closely related molecules undergoing the same irradiation process.

The ability of the RNN ensemble to be applied to different target molecules depends on the inclusion of the target molecules in the training set.
Models trained with data from both CH$_4$ and CHCl$_3$ targets performed similarly between the two.
This indicates that a single model can be trained on a range of target molecules and still produce accurate predictions.
However, attempts to generalize to molecules not included in the training set resulted in less accurate predictions (see Fig. S3 in Supplemental Materials for details).
We leave open the possibility that a model trained on a larger variety of target molecules may better generalize to new target molecules.

This work only considered irradiation of four small molecules.
Using only small molecules allowed us to provide a proof-of-concept test at low computational cost.
The method employed, though, is general and may be applied to a wide variety of academically and technologically relevant systems, e.g. nanowires, quantum dots, or photocatalysis. 
Computational investigation of irradiation processes in larger systems has been limited due to the cost required to sample adequately the possible interactions between the target material and the incident ions.
Wide ranges of incident ion energies, relative trajectories, and impact parameters must be considered.
Using an active learning approach to select new parameters, RNN ensembles can be trained from a limited number of simulations then used to predict the dynamics with different conditions at little computational expense.

The RNN ensemble method is also not limited to only the two target quantities examined in this paper.
Our provided code is able to select individual molecular orbitals, specified by orbital number, HOMO, or LUMO, and carry out predictions of change in occupation number, thereby enabling study of the occupation dynamics of individual orbitals of interest.
Alternatively, the provided code is readily adapted to consider the total occupation of all molecular orbitals.
Such a model would then predict the time-dependent ionization of the target material.
Our approach is, in principle, broadly icable to diverse time-dependent properties and materials systems.
Combining the RNN ensemble with active learning allows the machine learning models to accurately and consistently predict time-dependent processes using a minimal number of rt-TDDFT calculations.

\section*{Credit Statement}
\textbf{Ethan P. Shapera}: Conceptualization (equal), Investigation (equal), Data Curation (equal), Writing (Original Draft). 
\textbf{Cheng-Wei Lee}: Conceptualization (equal), Investigation (equal), Data Curation (equal), Writing (Original Draft). 

\section*{Acknowledgments}
Fruitful discussions with Andr\'{e} Schleife and Peter T. Leistikow are gratefully acknowledged. 
rt-TDDFT calculations were carried out using computational resources sponsored by the Department of Energy’s Office of Energy Efficiency and Renewable Energy, located at the National Renewable Energy Laboratory. 
The authors also acknowledge Colorado School of Mines supercomputing resources (http://ciarc.mines.edu/hpc) made available for conducting the research reported in this paper.
The views expressed in the article do not necessarily represent the views of the DOE or the U.S. Government.
Computational time for constructing machine learning models was provided in part by the dCluster of the Graz University of Technology.

\section*{Data Availability}
All raw rt-TDDFT data will be made available through the Materials Data Facility\cite{MDF,blaiszik2019}.

\subsection*{Code Availability}
Code to generate descriptors and perform the machine learning is available at Ref. \ \onlinecite{repo}.

\bibliography{./literature}

\clearpage

\appendix
\renewcommand{\thesubsection}{S\arabic{subsection}}
\renewcommand{\theequation}{S\arabic{equation}}
\renewcommand{\thetable}{S\arabic{table}}
\renewcommand{\thefigure}{S\arabic{figure}}
\setcounter{equation}{0}
\setcounter{table}{0}
\setcounter{figure}{0}
\section*{Supplemental Materials}

\FloatBarrier

\subsection{\label{subsec:TDDFTMethods} rt-TDDFT Methods}

We generated CH$_x$Cl$_y$ ($x$+$y$=4) molecule models using first-principles molecular dynamics with Qb@ll\cite{schleife:2012,draeger:2017} to break the intrinsic symmetry. 
The distorted structures prevent degenerate eigenstates and make each of the equally spaced proton trajectories unique. 
The molecular dynamics was simulated with time step of 10 a.u. ($\approx$0.24 fs) at 1000 K for 3000 ionic steps. 
The last molecular structures were used for the molecular configuration in the subsequent real-time simulations of proton irradiation.
The tetragonal simulation boxes were at least 30 Bohr radii (a$_B$) ($\approx$ 1,600 pm) in each lattice direction to model isolated molecules. 
Calculations were performed with a planewave cutoff energy of 100 Ry and a $\Gamma$-only $k$-mesh.  
Interactions were simulated by an (adiabatic) PBE exchange-correlation functional used with the electron-ion interactions described by the norm-conserving pseudopotentials as implemented by Hamann, Schl{\"u}ter, and Chiang pseudopotentials and modified by Vanderbilt \cite{Vanderbilt:1985}.
4, 1, and 7 valence electrons were considered for the pseudopotentials of C, H, and Cl, respectively. 

\begin{figure}[h!]
\includegraphics[width=0.9\linewidth]{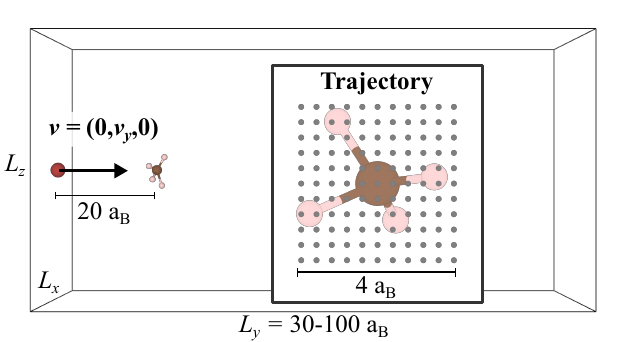}
\caption{\label{fig:impact}
Simulation setup for proton-molecule interactions. The inset shows the proton trajectories we sample for CH$_4$, viewed along the $y$-axis. The box length in y direction depends on the proton velocity and ranges from 30 a$\mathrm{_{B}}$ to 100 a$\mathrm{_{B}}$.
}
\end{figure}

Figure\ \ref{fig:impact} shows the setup to simulate the proton-molecule interactions. 
The proton was initially set 20 Bohr radii in the y-direction (a$\mathrm{_{B}}$) away from the C atom at simulation time 0.
In order to simulate different impact parameters, we selected a grid of starting proton initial positions in the XZ-plane. 
The grid points were equally spaced with 11 points in each direction with a grid length of 4 a$\mathrm{_{B}}$ for CH$_4$ or 6 a$\mathrm{_{B}}$ for Cl-containing molecules. 
The proton was set to travel along the y-direction with a range of fixed speeds.

To ensure there was sufficient time for electron dynamics to equilibrate after the proton-target interaction (shown in Figures\ \ref{fig:activeMAE-occ}c and \ref{fig:activeMAE-energy}c as the long-time oscillations), we adjusted the simulation box length in y-direction such that the proton reached the opposite end of the simulation box after at least 32 a.u. ($\approx$0.77 fs). 

The electron dynamics were propagated in real time using the enforced time-reversal symmetry (ETRS) method\cite{castro:2004,draeger:2017} with the time step of 0.04 a.u. ($\approx$0.967 as).
The chosen time step ensured numerical stability and charge conservation. 
The ionic positions were fixed during the molecule-proton interactions  to consider only the electron dynamics.
 
We monitored the electron dynamics during the molecule-proton interactions by energy gain and occupation number of each molecular orbital.
The occupation was calculated based on the time-dependent occupation number of eigenstate $i$, $f_{i}(t)$,
\begin{equation}
\label{eq:occ}
f_{i}(t)=\sum_{j=1}\left|\left\langle\phi_{i}|\psi_{j}(t)\right\rangle\right|^{2},
\end{equation}
where the reference states $\phi_{i}$ were the initial DFT-calculated molecular orbital eigenstates and $\psi_{j}(t)$ were the time-dependent molecular orbitals evolved from the initial eigenstates $j$ at time $t$.
The change in total occupation of the molecular valence orbital states ($\Delta O$) was then defined,
\begin{equation}
\label{eq:delo}
\Delta O=\sum_{occupied}f_i(t)-f_i(t=0),
\end{equation}
where $f_i(t=0)$ were the occupations of the initial orbital eigenstates at $t=0$ and only initially occupied states are considered.

\subsection{\label{subsec:EnergyModel}Energy Model}
\begin{figure}[htb!]
\includegraphics[width=0.8\linewidth]{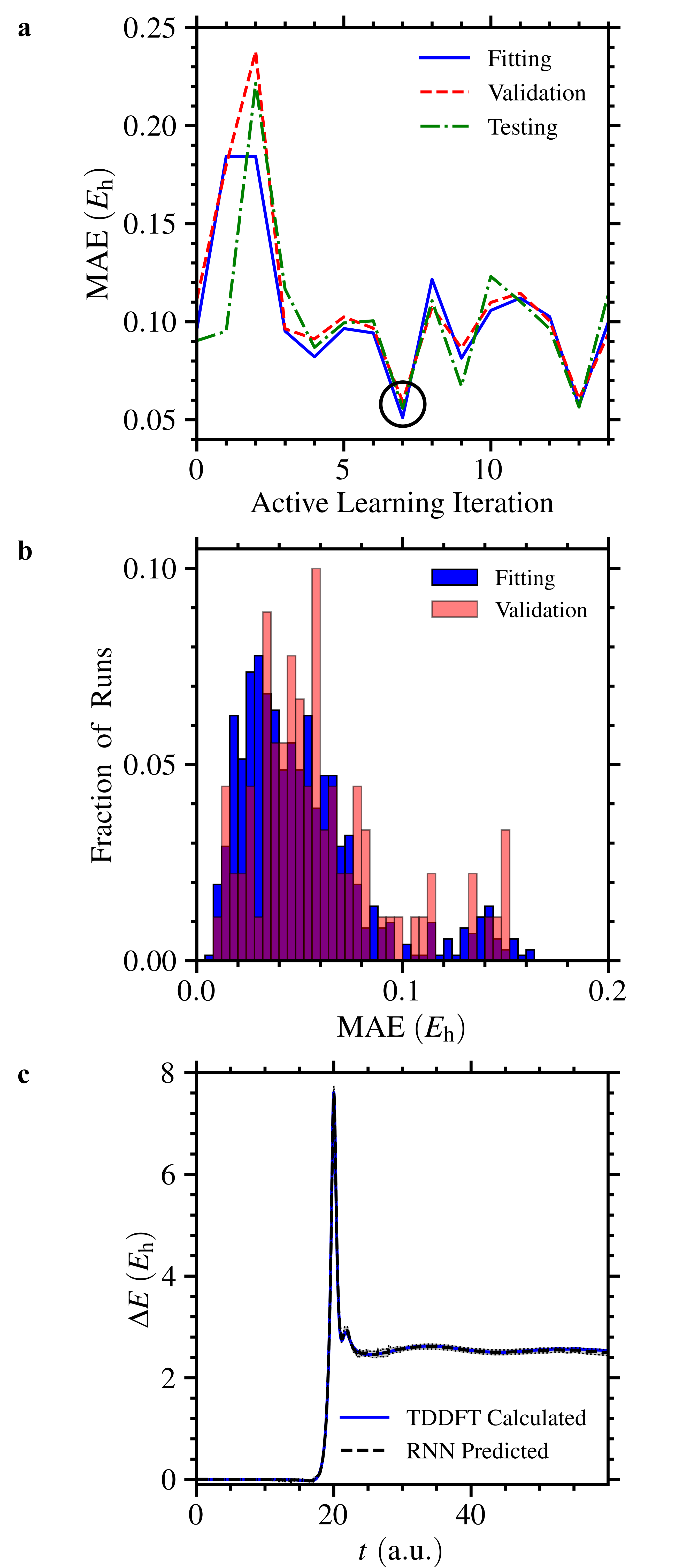}
\caption{\label{fig:activeMAE-energy}
a. Variation in MAE of total energy for ensemble of 10 RNNs trained on CH$_4$ and CHCl$_3$ as more data is added via active learning.
The seventh iteration, which minimizes the MAEs and is discussed further is circled.
b. Distribution of run MAE values in the seventh active learning iteration for all runs in the fitting and validation sets for the ensemble of 10 RNNs.
c.Representative example of RNN ensemble prediction results for the change in total energy.
A proton projectile with a speed of 1.0 a.u. passes within 0.4 a$_B$ ($\approx$ 21 pm) of the C-atom in CH$_4$.
Gray shaded regions show the 1$\sigma$ uncertainty of the RNN ensemble predictions.
}
\end{figure}

Figure \ \ref{fig:activeMAE-energy}a plots the MAE values of an RNN ensemble used to predict the change in total system energy over 15 iterations of active learning.
The MAE values for the fitting, validation, and testing sets decrease non-monotonically and reach a minimum at the seventh active learning iteration.
Similar to the occupation model, MAE values increase in each successive iteration after the seventh active learning iteration.

Figure \ \ref{fig:activeMAE-energy}b shows the distribution of MAE values for each run used in the fitting and validation sets for fitting each RNN in the seventh active learning iteration.
The seventh iteration is chosen because the MAE values reach a minimum at this iteration.
Here the MAE values are 0.051 $E_\mathrm{h}$($\approx$ 1.4 eV) for the fitting set, 0.059 $E_\mathrm{h}$for the validation set, and 0.056 $E_\mathrm{h}$for the testing set.
The matching MAE values for fitting and validation, along with the close match of the MAE distributions demonstrates that the model has not been overfit.

We provide a representative example of the performance of the RNN ensemble in the seventh active learning iteration for predicting the change in system energy during an irradiation process in Figure \ \ref{fig:activeMAE-energy}c.
In the representative example, a proton projectile is launched at a CH$_4$ molecule with a speed of 1.0 a.u.
The projectile reaches a minimum distance from the C-atom of 0.4 a$_B$ ($\approx$ 21 pm).
This specific run appeared in the reservoir set when training the RNN ensemble.
The RNN ensemble produced remarkable quantitative and qualitative agreement with the rt-TDDFT calculation.
The RNN ensemble has MAE of 0.012 $E_\mathrm{h}$averaged over the full run with a 1-$\sigma$ uncertainty of 0.04H, compared against a final change in energy of 2.53 $E_\mathrm{h}$.
As in the occupation model, the energy model correctly matches behavior in the short, medium, and long time regimes.

The RNN ensemble model developed here consistently provides reliable predictions of the dynamical behavior of small molecules subject to proton irradiation.
Models trained using a small number of speed-trajectory combinations for a given target molecule accurately predict the behavior for new projectile speed-trajectory combinations.

\subsection{\label{subsec:generalization}Molecular Target Generalization}

\begin{figure}[hbt!]
\includegraphics[width=0.7\linewidth]{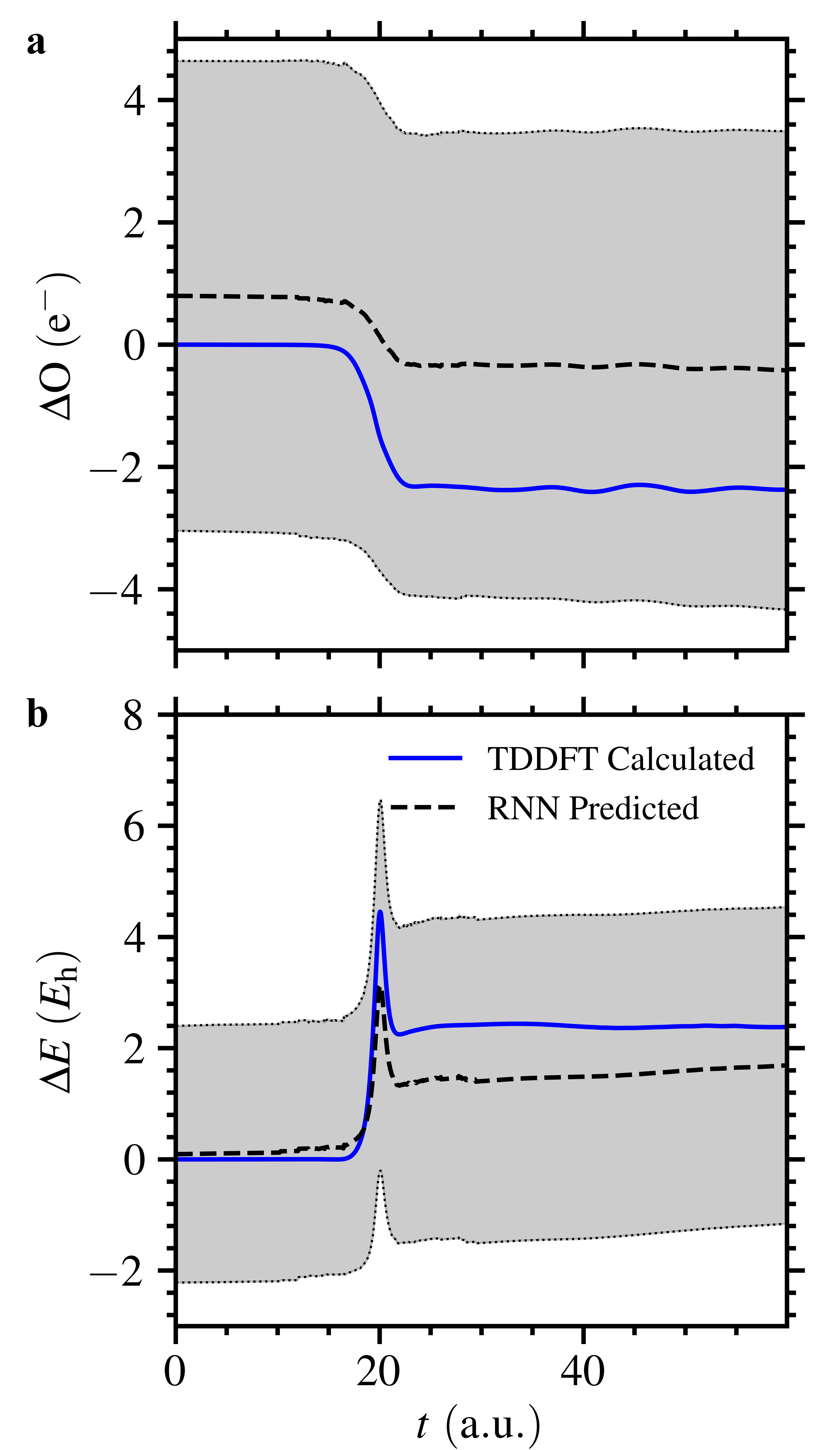}
\caption{\label{fig:ApplicationAttempt}
Comparison of TDDFT-calculated and RNN-predicted dynamics for a CH$_3$Cl subjected to irradiation by a proton. 
A proton with a speed of 1.0 a.u. passes within 0.6 a$_B$ ($\approx$ 32 pm) of the C-atom in the CH$_3$Cl target.
a. Change in total occupation of the valence orbitals.
b. Change in system energy.
}
\end{figure}

The RNN models trained using rt-TDDFT data for CH$_4$ and CHCl$_3$ targets was applied to predict the change in valence orbital occupation and projectile energy for CH$_3$Cl and CCl$_4$ target molecules when irradiated by a proton projectile.
Figure \ \ref{fig:ApplicationAttempt} provides a representative example of the RNN model performance if employed to predict the time-dependent behavior for a CH$_3$Cl target molecule.
In this representative example, a proton was launched at a CH$_3$Cl molecule with a speed of 1.0 a.u ($\approx$ 2.2 $\times$ 10$^6$ m s$^{-1}$).
The projectile reaches a minimum distance of 0.6 a$_B$ ($\approx$ 32 nm) from the central C atom.

Figure \ \ref{fig:ApplicationAttempt}a considers the change in valence orbital occupation.
The RNN ensemble prediction has an average MAE of 1.6 e$^-$ and 1-$\sigma$ uncertainty of 3.8 e$^-$ averaged over the irradiation process.
Figure \ \ref{fig:ApplicationAttempt}b compares the TDDFT-calculated system energy against the prediction made by the ensembles of RNNs.
The RNN ensemble produces a MAE of 0.65 $E_\mathrm{h}$with a 1-$\sigma$ uncertainty of 2.7 $E_\mathrm{h}$.
The RNN ensemble performs one to two orders of magnitude worse in both MAE and uncertainty when used to generalize to new target molecules than when predicting changes in energy for target molecule systems used to train the model.

Despite the poor quantitative agreement of the RNN ensembles for generalization, there are two key results to note.
First, while the RNN ensembles do not quantitatively match the rt-TDDFT calculations, they qualitatively reproduce several key features.
The models correctly identify that the valence occupation and energy are constant before the projectile interacts with the target molecule.
At intermediate times ($\approx$20 a.u. in the representative example), the RNN ensembles correctly identify the closest approach of the projectile to the target molecule.
At long times, the ensemble predicts slow, low amplitude oscillations.

The second notable behavior of the RNN ensemble predictions is the large uncertainties in predictions when generalization is attempted.
For both the occupation and energy models, the standard deviations are two orders of magnitude when used to predict quantities for target molecules not in the training data compared to predictions for target molecules included in the training data.
The large uncertainties in generalization are beneficial because they indicate that the RNN ensemble models are not hallucinating accurate results.

\end{document}